\documentclass[showpacs,amsmath,amssymb,preprint]{revtex4}
\usepackage{here}

\usepackage{setspace}
\usepackage{graphicx}% Include figure files
\usepackage{dcolumn}% Align table columns on decimal point
\usepackage{bm}% bold math
\usepackage{color}
\begin{document}

%\preprint{APS/123-QED}

\title{Particle-Localized Ground State of Atom-Molecule Bose-Einstein Condensates in a Double-Well Potential}% Force line breaks with \\

\author{Atsushi Motohashi}
% \altaffiliation[Also at ]{Physics Department, XYZ University.}%Lines break automatically or can be forced with \\
\author{Tetsuro Nikuni}%
%\email{Second.Author@institution.edu}
\affiliation{%
Department of Physics, Tokyo University of Science,
 1-3 Kagurazaka, Shinjuku-ku, Tokyo 162-8601, Japan\\
%This line break forced with \textbackslash\textbackslash
}%

\date{\today}% It is always \today, today,
             %  but any date may be explicitly specified

\begin{abstract}
We study the effect of atom-molecule internal tunneling on the ground state of atom-molecule Bose-Einstein condensates in a double-well potential.  In the absence of internal tunneling between atomic and molecular states, the ground state is symmetric, which has equal-particle populations in two wells. From the linear stability analysis, we show that the symmetric stationary state becomes dynamically unstable at a certain value of the atom-molecule internal tunneling strength. Above the critical value of the internal tunneling strength, the ground state bifurcates to the particle-localized ground states. The origin of this transition can be attributed to the effective attractive inter-atomic interaction induced by the atom-molecule internal tunneling. This effective interaction is similar to that familiar in the context of BCS-BEC crossover in a Fermi gas with Feshbach resonance. Furthermore, we point out the possibility of reentrant transition in the case of the large detuning between the atomic and molecular states.
\end{abstract}

\pacs{Valid PACS appear here}% PACS, the Physics and Astronomy
                             % Classification Scheme.
%\keywords{Suggested keywords}%Use showkeys class option if keyword
                              %display desired
\maketitle

\section{\label{sec:level1}Introduction}
Bose-Einstein condensation in dilute atomic gases has been offering opportunities to research macroscopic quantum phenomena, since its experimental realization in 1995. In particular, one of the most fascinating macroscopic quantum phenomena is the Josephson effect between two Bose-Einstein condensates(BECs) trapped in a double-well potential. This system is called as a Boson Josephson Junction(BJJ)\cite{Raghavan}. Recently BJJ has been realized experimentally, and macroscopic wavefunctions are observed directly\cite{BJJ}. This experimental achievement has triggered many interesting researches\cite{Chaiotic_BJJ,Shchesnovich_Trippenbach,Dissipated_BJJ}. Though in BJJ the spatial coherence of BECs is focussed, Josephson effects occur not only between spatial separated BECs but also between internal degrees of freedom in a single BEC. In particular, Josephson-like effects between atomic and molecular states have been discussed theoretically\cite{Rarified_Liquid_AM_BEC,Guang,Santos}.  

In the last decade many efforts have been devoted to creating molecular BECs from ultracold atoms\cite{Heinzen_Molecule,AM_BEC_Wieman,Fermion_Molecule_BEC,Feshbach_Mole_from_BEC,Aom_Molecule_dark_state}. Already molecular BECs have been created from fermionic atoms using magnetic Feshbach resonances\cite{Fermion_Molecule_BEC}. On the other hand, the creation of coexisting atomic and molecular condensates by means of photoassociation has been discussed theoretically\cite{Javanainen_PA_PRL,H_Pu_PA_PRL,Itin_Watanabe}. Photoassociation permits precise control of population transfer between individual discrete quantum states\cite{PARev}. Currently, a mixture of a Rb BEC and a degenerate gas of $\textrm{Rb}_{2}$ ground-state molecules has been realized using photoassociation\cite{Aom_Molecule_dark_state}. Furthermore, though not in Bose-Einstein condensed phase, the collective oscillation of the populations between an atomic state and a molecular state has been observed\cite{AM_BEC_Wieman,Ryu,AMRabi}. The realization of atom-molecule BECs is forthcoming, and these experimental achievements have accelerated many theoretical researches about atom-molecule coherence\cite{Molecular_Solitons,superchemistry,Anglin_Vardi_AM,Miyakawa_BF_coherence,M_Jack_Han_Pu,Miyakawa_dissociation}. In particular it is discussed that the atom-molecule internal tunneling changes the nature of system drastically. For instance it is predicted that atom-molecule internal tunneling can induce a droplet-like ground state in atom-molecule BEC mixtures\cite{Rarified_Liquid_AM_BEC}. The relation between the Ising model and the phase transition of bosonic atom-molecule mixtures is also discussed\cite{AM_Rad_2004,AM_Stoof_2004,AM_Bhaseen_2009}. As for atom-molecule mixtures in optical lattices, the possibility of so-called ``super-Mott" phase is pointed out\cite{AM_Mixture_OL,AM_Mixture_OL_PRL}.

%\cite{two_component_double_well}\cite{BJJ_chaos_Bunary}\cite{Binary_BJJ}\cite{Ashhab_Lobo}\cite{SB_two_BEC_DW}\cite{Motohashi_P1}
In this paper, we study atom-molecule BECs in a double-well potential by focusing on the effect of atom-molecule internal tunneling on the ground state. Although several authors have discussed BJJ of binary mixtures\cite{Ashhab_Lobo,two_component_double_well,SB_two_BEC_DW,Binary_BJJ}, effects of internal degrees of freedom in BECs in a double-well potential have not been fully discussed. In the present paper, we consider the atom-molecule internal tunneling. Even in a single-component BJJ, the competition between strengths of tunneling and interaction causes various phenomena such as macroscopic quantum self-trapping(MQST)\cite{Raghavan}. Adding an atom-molecule internal tunneling as a new degree of freedom, we will show that the competition between the atom-molecule internal tunneling and inter-well tunneling or interaction leads to new phenomena.

As our main result, we will show that atom-molecule internal tunneling induces the asymmetric ground state, which has non-equal particle populations in two wells. In the absence of internal tunneling, the ground state is symmetric, with equal particle populations in two wells. We note that ground states breaking the symmetry of trapping potentials have been found in various BEC systems. The well-known example of a symmetry-breaking ground state is a soliton in a quasi-one-dimensional attractive BEC\cite{Soliton_Ueda_PRL,Trippenbach_QPT_Attractive}. The ground state in attractive BJJ also breaks the left-right symmetry of the double-well potential above certain value of the interaction strength as predicted theoretically\cite{Trippenbach_QPT_BJJ}. The asymmetric ground states in the above systems are caused by attractive interactions. In contrast we will show that, even for repulsively interacting Bose gasses, spontaneous symmetry breaking in the ground state emerges in atom-molecule BECs in a double-well potential owing to the atom-molecule internal tunneling. 

In addition, we show in the simplest case that the effect of atom-molecule internal tunneling can be described in terms of the effective inter-atomic attractive interaction. This effective interaction is similar to that familiar in the context of BCS-BEC crossover in a Fermi gas with Feshbach resonance\cite{Ohashi_Griffin_PRL}. We show that this effective interaction is always attractive and induces the asymmetric ground states in the absence of the molecular tunneling, the inter-molecule interaction and the inter atom-molecule interaction. Furthermore, we discuss the possibility of the reentrant transition, which cannot be explained by the simple form of the effective attractive interaction.

This paper is organized as follows. In Sec. \ref{sec:model}, we explain the model and approximations, used in this paper. In Sec. \ref{sec:model_approximation}, we introduce a four-mode model and classical analysis. In the four-mode model, we concentrate on condensate modes only and ignore other modes. Furthermore, we ignore quantum fluctuations by replacing creation-annihilation operators in Hamiltonian by c-number. Next, the parameters in this model are estimated from experiments. 

In order to investigate the ground state, we first derive time-evolution equations in Sec. \ref{sec:eq_time_evolution}. Second, by using these equations, we derive the equations for the particle populations in the ground states in Sec. \ref{sec:eq_ground}, and we develop the expression for eigen frequencies in Sec. \ref{sec:excitation_spectra}. 

In Sec. \ref{sec:symmetry-breaking_ground_states}, we perform the linear stability analysis, using the equations derived in Sec \ref{sec:model}. In particular, we investigate the stability of symmetric stationary states, where the particle numbers in the left and right wells are equal. By performing the linear stability analysis, we show that the atom-molecule tunneling induces the dynamical instability of the symmetric stationary state, which is the ground state in the absence of the atom-molecule tunneling. This indicates the emergence of the symmetry breaking ground states, where the particles localize in one well. In Sec. \ref{sec:localization}, by using the equations for the particle populations in the ground state derived in Sec. \ref{sec:eq_ground}, we show this instability is accompanied with the bifurcation of symmetric stationary states to asymmetric ones. By comparing the energies of symmetric and asymmetric states, we confirm that the asymmetry state is the ground state. The general relation between dynamical instability and  phase transition of ground states is discussed briefly in appendix \ref{sec:effective_potential}. Some details of the calculations are given in Appendices \ref{sec:parameter_appendix} - \ref{sec:app_1}.

\section{\label{sec:level1}Model and Approximations}
\label{sec:model}
\subsection{Four-mode model and classical analysis}
\label{sec:model_approximation}
The second-quantized Hamiltonian for Bose atoms and molecules can be written as
\begin{eqnarray}
&& \hat{H} = \sum_{i=a,b} \int d \mathbf{r} \Bigg( \frac{ \hbar^2 }{ 2m_{i} } \nabla \hat{ \Psi }_{i}^{ \dag } \cdot \nabla \hat{ \Psi }_{i} + V_{ \rm{ext} } ( \mathbf{ r }  )  \hat{ \Psi }_{i}^{ \dag } \hat{ \Psi }_{i} \Bigg) \nonumber \\
&& \qquad + \frac{g_{i}}{2} \sum_{ i=a,b } \int d \mathbf{r} \hat{ \Psi }_{i}^{ \dag } \hat{ \Psi }_{i}^{ \dag } \hat{ \Psi }_{i} \hat{ \Psi }_{i} + g_{ab} \int d \mathbf{r} \hat{ \Psi }_{a}^{ \dag } \hat{ \Psi }_{b}^{ \dag } \hat{ \Psi }_{b} \hat{ \Psi }_{a} \nonumber \\
&& \qquad - \lambda \int d \mathbf{r} \left( \hat{ \Psi }_{b}^{ \dag } \hat{ \Psi }_{a} \hat{ \Psi }_{a} + \hat{ \Psi }_{a}^{\dag}  \hat{ \Psi }_{a}^{\dag} \hat{ \Psi }_{b} \right) + \delta \int d \mathbf{r} \hat{ \Psi }_{b}^{ \dag } \hat{ \Psi }_{b} ,  \qquad 
\end{eqnarray}
where $\hat{ \Psi }_{a}$ and $\hat{ \Psi }_{b}$ represent  field operators for Bose atoms and molecules respectively, $\lambda$ is the internal tunneling strength between atomic and molecular states, $\delta$ is the energy difference between atoms and molecules, and $V_{ \rm{ext} } ( \mathbf{ r } )$ is a double-well potential. The inter-atomic, the inter-molecule, and the atom-molecule interactions can be approximated in terms of the s-wave scattering lengths as $g_{i} = 4 \pi \hbar^2 a_{s i} / m_{i}, g_{ab} = 6 \pi \hbar^2 a_{s ab} / m_{a}({\it i}=a,b ,m_{b}=2m_{a})$. Here, $m_{a}$ is the mass of a Bose atom. Furthermore, we introduce the four-mode approximation. In this approximation, we concentrate on condensate modes only, and ignore the effect of the particles occupying other modes. From this point of view, field operators can be approximated as $\hat{ \Psi }_{a} \simeq \Phi_{aL} \hat{a}_{L} + \Phi_{aR} \hat{a}_{R}, \hat{ \Psi }_{b} \simeq \Phi_{bL} \hat{b}_{L} + \Phi_{bR} \hat{b}_{R}$, where $\Phi_{aL}, \Phi_{aR}$($\Phi_{bL}, \Phi_{bR}$) are the wavefunctions of the atomic(molecular) condensate modes in the left well and the right well respectively. $\hat{a}_{L}, \hat{a}_{R}$($\hat{b}_{L}, \hat{b}_{R}$) are annihilation operators for the atomic(molecular) condensate modes in the left well and the right well respectively. Applying these approximations to Eq.(1), we obtain the quantum four-mode Hamiltonian (four-mode model). 
\begin{eqnarray}
\hat{H} &=& - J_{a} \big( a_{L}^{\dag} a_{R} +  a_{R}^{\dag} a_{L} \big) - J_{b} \big( b_{L}^{\dag} b_{R} + b_{R}^{\dag} b_{L} \big) + \Delta \big( b_{L}^{\dag} b_{L} + b_{R}^{\dag} b_{R} \big)  \nonumber \\
&& + \frac{ U_{a} }{2} \big( a_{L}^{\dag} a_{L}^{\dag} a_{L} a_{L} + a_{R}^{\dag} a_{R}^{\dag} a_{R} a_{R} \big) + \frac{ U_{b} }{2} \big( b_{L}^{\dag} b_{L}^{\dag} b_{L} b_{L} + b_{R}^{\dag} b_{R}^{\dag} b_{R} b_{R} \big) \nonumber \\
&& + U_{ab} \big( a_{L}^{\dag} a_{L} b_{L}^{\dag} b_{L} + a_{R}^{\dag} a_{R} b_{R}^{\dag} b_{R} \big) - g \big( b_{L}^{\dag} a_{L} a_{L} + b_{R}^{\dag} a_{R} a_{R} + a_{L}^{\dag} a_{L}^{\dag} b_{L} + a_{R}^{\dag} a_{R}^{\dag} b_{R} \big) , \quad
\label{eq:am_H}
\end{eqnarray}
where the parameters are defined in Appendix \ref{sec:parameter_appendix}. In addition, we use classical analysis, in which annihilation operators are replaced by c-number $ \sqrt{ N } e^{ i \theta } $, where {\it N} is the particle number of a condensate mode and $\theta$ is its phase. This approximation is justified when an occupation number is macroscopic. Using this procedure, we obtain the classical four-mode Hamiltonian as
\begin{eqnarray}
&& H_{cl} = - 2 J_{a} \sqrt{N_{aL}N_{aR}} \cos ( \theta_{aR} - \theta_{aL} ) - 2 J_{b} \sqrt{N_{bL}N_{bR}} \cos ( \theta_{bR} - \theta_{bL}  ) + \Delta ( N_{bL} + N_{bR} ) \nonumber \\
&& \qquad \quad + \frac{U_{a}}{2} ( N_{aL}^2 + N_{aR}^2 ) + \frac{U_{b}}{2} ( N_{bL}^2 + N_{bR}^2 ) + U_{ab} ( N_{aL} N_{bL} + N_{aR} N_{bR} ) \nonumber \\
&& \qquad \quad - 2g \big[ N_{aL} \sqrt{N_{bL}} \cos ( 2 \theta_{aL} - \theta_{bL} ) + N_{aR} \sqrt{ N_{bR} } \cos ( 2 \theta_{aR} - \theta_{bR}  ) \big],
\label{eq:classical_Hamiltonian}
\end{eqnarray}
where $N_{aL(aR)}$ represents the particle number of the atom in the left(right) well, and $N_{bL(bR)}$ does that of the molecule in the left(right) well. $\theta_{aL(aR)}$ is the phase of the atomic condensate in the left(right) well, and $\theta_{bL(bR)}$ is the phase of the molecular condensate in the left(right) well.

%\subsubsection{Experimental parameters}
%\label{sec:parameter}
In order to relate our model to realistic systems, we consider the double-well trap potential used in the experiment of a single-component BJJ of $^{87}\textrm{Rb}$\cite{BJJ}, and set the parameters to be consistent with this experiment. In this experiment, parameters are as follows. The ratio $\Lambda = N U_{a} / ( 2 J_{a} )$ is estimated as $15$ in Ref.\cite{BJJ}, which corresponds to the strong-coupling case. Since the total-particle number $N$ is $1150$ in \cite{BJJ}, the atomic interaction strength normalized by the atomic tunneling strength can be obtained as $U_{a} / J_{a} \simeq 3 \times 10^{-2}$. We use this value for the atomic interaction strength. As for the molecular interaction strength, we suppose that the molecular scattering length is the same as the atomic one and that the shape of condensate wavefunctions of atoms and molecules are the same. Under this condition $U_{b} = U_{a} / 2$ from Eq. (\ref{eq:p_4}). In addition, in this study we set the total particle number as ${\it N} = {\it N}_{aL} + {\it N}_{aR} + 2 {\it N}_{bL} + 2 {\it N}_{bR} = 2000$.

We next consider the atom-molecule interaction. From the experiment\cite{Heinzen_Molecule}, the atom-molecule scattering length of $^{87} \textrm{Rb}$ is estimated as $a_{am} = - 180 \pm 150 a_{0}$, where $a_{0}$ is Bohr radius, and the ratio of the atom-molecule scattering length and the atomic scattering length can be derived as $a_{am} / a_{a} \simeq - 3.2 \sim 0.3$. Based on this value we suppose that the atom-molecule interaction $U_{ab}$ to be negative, but treat it as a variable parameter. We will compare results with different values of $U_{ab}$. The negative $U_{ab}$ is well-suited to the internal tunneling between internal states because phase separation does not occur. This is different from a binary BEC mixture in the $| F = 2, m_{f} = 2 \rangle$ and $| 1, - 1 \rangle$ spin states of $^{87} \textrm{Rb}$, where component separation is observed due to the repulsive interspecies interaction\cite{Hall_separate_binary_BEC}. 

In order to set the molecular tunneling strength, we suppose that the atomic eigen state $\Phi_{aL(aR)}$ and the molecular eigen state $\Phi_{bL(bR)}$ have almost the same shapes. From the Eq. (\ref{eq:p_1}) and $m_{b} = 2 m_{a}$, $J_{b} / J_{a}= 1/2$.

%Finally we remark that the four-mode model is a one-dimensional model. Within this model, in respect to spatial degree of freedoms, we treat only the direction in which inter-well Josephson oscillations occur.

\subsection{Equations of time evolution}

\label{sec:eq_time_evolution}

In order to obtain the ground state and the eigen frequencies corresponding to the excitation spectra, we derive the time-evolution equations. Using the classical four-mode Hamiltonian, we can derive the Hamilton equations of motion describing the dynamics of atomic and molecular BECs in a double-well potential as :
\begin{eqnarray}
&& \hbar \dot{N}_{aL} = \frac{ \partial H_{cl} }{ \partial \theta_{aL} } , \quad \hbar \dot{N}_{aR} = \frac{ \partial H_{cl} }{ \partial \theta_{aR} }, 
\label{eq:canonical_1}\\
&& \hbar \dot{N}_{bL} = \frac{ \partial H_{cl} }{ \partial \theta_{bL} } , \quad \hbar \dot{N}_{bR} = \frac{ \partial H_{cl} }{ \partial \theta_{bR} },
\label{eq:canonical_2} \\ 
&& \hbar \dot{ \theta }_{aL} = - \frac{ \partial H_{cl} }{ \partial N_{aL} } , \quad \hbar \dot{ \theta }_{aR} = - \frac{ \partial H_{cl} }{ \partial N_{aR} }, 
\label{eq:canonical_3} \\
&& \hbar \dot{ \theta }_{bL} = - \frac{ \partial H_{cl} }{ \partial N_{bL} } , \quad \hbar \dot{ \theta }_{bR} = - \frac{ \partial H_{cl} }{ \partial N_{bR} }.
\label{eq:canonical_4}
\end{eqnarray}
These time evolution equations can also be obtained by applying classical analysis to the Heisenberg equations derived from the quantum four-mode Hamiltonian. Heisenberg equation and Hamilton equation are related through the canonical commutation relation $\left[ \hbar \hat{N} , \hat{ \theta } \right] = i \hbar$ and the Poisson bracket $\{ \hbar N , \theta \} = 1 $. The explicit expression of these equations are given in Appendix \ref{sec:app_1_1}. 
 
\subsection{Equations for ground state}
\label{sec:eq_ground} 
We now look for stationary solutions of the equations of motion for the particle numbers (\ref{eq:te_1}) - (\ref{eq:te_7}). We can easily find that the relative phases should be $0$ in the ground states i.e. the atomic relative phase $\theta_{aL} - \theta_{aR} = 0$, the molecular relative phase $\theta_{bL} - \theta_{bR} = 0$, and the atom-molecule relative phases $2 \theta_{aL(aR)} - \theta_{bL(bR)} = 0$, respectively.

In this study, we investigate the ground states in the presence of the atom-molecule internal tunneling in a symmetric double-well potential. From the Hamiltonian (\ref{eq:classical_Hamiltonian}), the competition between the tunneling strengths and the interparticle interactions determines the ground states. The inter-well tunnelings $J_{a}, J_{b}$ lower the energy most when the particle populations are equal in two wells. The inter-atomic and inter-molecular repulsive interactions $U_{a}, U_{b}$ act in the same way. In contrast to the above contributions, the attractive atom-molecule interaction $U_{ab}$ lowers the energy when the particles localize in one well. As discussed later in Sec. \ref{sec:effective_attractive}, the atom-molecule internal tunneling $g$ can act effectively as the attractive inter-atomic interaction. This competition will cause the asymmetric ground state, which breaks the symmetry of the double-well trap potential.

In order to look for the ground state with a fixed particle number, we introduce the chemical potential $\mu$ and the grand canonical energy
\begin{eqnarray}
K \equiv H_{cl} - \mu N ,
\end{eqnarray}
where the total number is defined as $N = N_{aL} + N_{aR} + 2 N_{bL} + 2 N_{bR}$. Then the ground state can be determined from
\begin{eqnarray}
\frac{ \partial K }{ \partial N_{aL(aR)} } = 0, \quad \frac{ \partial K }{ \partial N_{bL(bR)} } = 0,
\label{eq:grand_ground}
\end{eqnarray}
or equivalently ( from Eqs. (\ref{eq:canonical_3}) and (\ref{eq:canonical_4}) )
\begin{eqnarray}
\hbar \dot{ \theta }_{aL(aR)} = - \mu , \quad \hbar \dot{ \theta }_{bL(bR)} = - 2 \mu .
\end{eqnarray}
From Eq. (\ref{eq:te_2})-(\ref{eq:te_8}), we obtain
\begin{eqnarray}
- \mu &=& J_{a} \sqrt{ \frac{ N_{aR} }{ N_{aL} } }  - N_{aL} U_{a} - N_{bL} U_{ab} + 2 g \sqrt{N_{bL} }, \label{eq:stationary_1} \\
- \mu &=& J_{a} \sqrt{ \frac{ N_{aL} }{ N_{aR} } } - N_{aR} U_{a} - N_{bR} U_{ab} + 2 g \sqrt{N_{bR} }, \label{eq:stationary_2} \\
- 2 \mu &=& J_{b} \sqrt{ \frac{ N_{bR} }{ N_{bL} } } - \Delta - U_{b} N_{bL} - N_{aL} U_{ab} + g \frac{ N_{aL} }{ \sqrt{ N_{bL} } },  \label{eq:stationary_3} \\
- 2 \mu &=& J_{b} \sqrt{ \frac{ N_{bL} }{ N_{bR} } }  - \Delta - U_{b} N_{bR} - N_{aR} U_{ab} + g \frac{ N_{aR} }{ \sqrt{ N_{bR} } }.  \label{eq:stationary_4} 
\end{eqnarray}
These equations determine the stationary states. It is clear that the above equations always have a symmetric solution $N_{aL} = N_{aR}, N_{bL} = N_{bR}$. However, the symmetric solution does not always have the lowest energy. We will show this in Sec. \ref{sec:symmetry-breaking_ground_states} from both the linear stability analysis and calculating energy.

\subsection{Excitation spectra}
\label{sec:excitation_spectra}
Solving the linearized Hamilton equations, we obtain the four eigen-frequencies of the excitation spectra from the stationary states. We will look at these eigen frequencies to investigate the stability of the system. If the excitation frequency $\omega$ has an imaginary part, such stationary state is dynamically unstable, i.e. if the stationary state is perturbed slightly, the small-amplitude oscillation exponentially grows in time.

Our procedure is summarized as follows. We first expand $H_{cl}$ in fluctuations around the symmetric stationary state to second order. Next, by performing the canonical transformation$\left( \theta_{aL}, \theta_{aR}, \theta_{bL}, \theta_{bR} \right) \rightarrow \left( \tilde{ \phi }_{0}, \tilde{ \phi }_{AM}, \tilde{ \phi }_{+}, \tilde{ \phi }_{-} \right)$, and $\left( {\it N}_{aL}, {\it N}_{aR}, {\it N}_{bL}, {\it N}_{bR} \right) \rightarrow \left( \tilde{ X }_{0}, \tilde{ X }_{AM}, \tilde{ X }_{+}, \tilde{ X }_{-} \right)$, we diagonalize the canonical momentum part of  the Hamiltonian. Details of calculations are given in Appendix \ref{sec:canonical_variables}. We arrive at the quadratic Hamiltonian
\begin{eqnarray}
H_{cl} &\simeq& \frac{1}{2} \tilde{ \phi }_{0}^{2} + \frac{ 1 }{ 2 } \tilde{ \phi }_{AM}^2 + \frac{ 1 }{ 2 } \tilde{ \phi }_{+}^2 + \frac{ 1 }{ 2 } \tilde{ \phi }_{-}^2 + V_{eff} \left( \tilde{ X }_{+}, \tilde{ X }_{-}, \tilde{ X }_{AM} \right).
\label{eq:di_Ham}
\end{eqnarray}
The explicit form of $V_{eff}$ is given by 
\begin{eqnarray}
V_{eff} \left( \tilde{ X }_{+}, \tilde{ X }_{-}, \tilde{ X }_{AM} \right) &\simeq& g N_{a} \sqrt{ N_{b} } \left[  4 U_{a} + U_{b} - 4 U_{ab} + \frac{ g }{ \sqrt{ N_{b} } } \left( 4 + \frac{ N_{a} }{ 2 N_{b} } \right) \right] \left( \delta \tilde{ X }_{AM} \right)^2 \nonumber \\
&& + \left( \delta \tilde{ X }_{+} , \delta \tilde{ X }_{-} \right) {\mathbf V } \left(
\begin{array}{cccc}   
\delta \tilde{ X }_{+} \\ 
\delta \tilde{ X }_{-} \\ 
\end{array}
\right), 
\label{eq:V_eff_a_0}
\end{eqnarray}
where $N_{a}\left(N_{b}\right)$ is the atomic(molecular) particle number in each well at the symmetric stationary state, that is, $N_{a} \equiv N_{aL(aR)}, N_{b} \equiv N_{bL(bR)}$. The $2 \times 2$ matrix ${\mathbf V}$ is defined as
\begin{eqnarray}
{\mathbf V } \equiv \left(
\begin{array}{cccc} 
2 \Omega_{+} Z_{+} \left( \alpha_{+}^2 J_{a}^{e} + J_{b}^{e} + 2 \alpha_{+} U_{ab}^{e} \right) , 2 \sqrt{ \Omega_{+} \Omega_{-} Z_{+} Z_{-} } \left( - J_{a}^{e} + J_{b}^{e} + \left( \alpha_{+} + \alpha_{-} \right) U_{ab}^{e} \right) \\ 
2 \sqrt{ \Omega_{+} \Omega_{-} Z_{+} Z_{-} } \left( - J_{a}^{e} + J_{b}^{e} + \left( \alpha_{+} + \alpha_{-} \right) U_{ab}^{e} \right) , 2 \Omega_{-} Z_{-} \left( \alpha_{-}^2 J_{a}^{e} + J_{b}^{e} + 2 \alpha_{-} U_{ab}^{e} \right)
\end{array} 
\right),
\end{eqnarray}
where $J_{a}^{e} \equiv J_{a} / N_{a} + U_{a}, J_{b}^{e} \equiv J_{b} / N_{b} + U_{b}^{e}$ and
\begin{eqnarray}
U_{b}^{e} \equiv U_{b} + \frac{ g N_{a} }{ 2 N_{b} \sqrt{ N_{b} } } , \quad U_{ab}^{e} \equiv \frac{ g }{ \sqrt{ N_{b} } } - U_{ab}.
\end{eqnarray}
Using this quadratic Hamltonian, we obtain the linearized Hamilton equations. Eliminating the phase variables, we arrive at 
\begin{widetext} 
\begin{eqnarray}
\hbar^2 \delta \ddot{ \tilde{X} }_{AM} &=& - 2 g N_{a} \sqrt{ N_{b} } \left( 4 U_{a} + U_{b}^{e} + 4 U_{ab}^{e} \right) \delta \tilde{ X }_{AM}, \label{eq:eigen_AM_19} \\
\hbar^2 \frac{ d^2 }{ d t^2  } 
\left(
\begin{array}{cccc} 
 \delta \tilde{ X }_{+} \\ 
 \delta \tilde{ X }_{-} \\   
\end{array} 
\right)
&=&  
- 2 {\mathbf V}
\left(
\begin{array}{cccc}   
 \delta \tilde{ X }_{+} \\ 
 \delta \tilde{ X }_{-} 
 \end{array}
 \right).
\label{eq:symmeric_linear}
\end{eqnarray}
\end{widetext}
We then assume the normal mode solutions, $\delta \tilde{ X }_{\pm} \propto e^{\pm i \omega t}, \delta \tilde{ X }_{AM} \propto e^{\pm i \omega t}$. Equation (\ref{eq:eigen_AM_19}) immediately gives one eigen frequency:
\begin{eqnarray}
\left( \hbar \omega_{ AM } \right)^2 = 2 g N_{a} \sqrt{ N_{b} } \left( 4 U_{a} + U_{b}^{e} + 4 U_{ab}^{e} \right)
\end{eqnarray}
The other two eigen frequencies can be obtained by diagonalizing the coefficient matrix of Eq. (\ref{eq:symmeric_linear}):
%\begin{widetext} 
\begin{eqnarray}
\left( \hbar \omega_{ \pm } \right)^2 &=& 2 \left( A J_{a}^{e} + C J_{b}^{e} \right) + 4 B U_{ab}^{e} \nonumber \\
&& \pm \sqrt{ 4 \left( \left( A J_{a}^{e} + C J_{b}^{e} \right) + 2 B U_{ab}^{e} \right)^2 - 16 \left( A C - { B }^{2} \right) \left( J_{a}^{e} J_{b}^{e} - \left( U_{ab}^{e} \right)^2 \right) }, 
\label{eq:spectra_new}
\end{eqnarray}
%\end{widetext}
where we have defined the coefficients {\it A}, {\it B} and {\it C} as $A \equiv N_{a} J_{a} + 2 g N_{a} \sqrt{ N_{b} }, B \equiv g N_{a} \sqrt{ N_{b} }$ and $C \equiv N_{b} J_{b} + \frac{ 1 }{ 2 } g N_{a} \sqrt{ N_{b} }$ respectively. We note that from Eq. (\ref{eq:ryusi_mode}), $\tilde{X}_{AM}$ and $\tilde{X}_{\pm}$ can be expressed in terms of number variables as
\begin{eqnarray}
\tilde{ X }_{AM} &=& \frac{N_{b} - 2 N_{a}}{10 \sqrt{ g N_{a} \sqrt{ N_{b} } } }. 
\label{eq:Xam} \\
\tilde{ X }_{\pm} &=& \pm \frac{ \left[ \left( N_{aL} - N_{aR} \right) + \alpha_{\mp} \left( N_{bL} - N_{bR} \right) \right] }{ 2 \left( \alpha_{+} - \alpha_{-} \right) \sqrt{ 2 \Omega_{\pm} Z_{\pm} } }, 
\label{eq:Xpm} 
\end{eqnarray}
From Eq. (\ref{eq:Xam}), we see that $\tilde{X}_{AM}$ mode represents the oscillation between atomic and molecular BECs. Therefore, $\omega_{AM}$ represents the internal Josephson frequency between atomic-molecular states. Since one can easily show $\alpha_{+} > 0$ and $\alpha_{-} < 0$, Eq. (\ref{eq:Xpm}) clearly shows that $\tilde{ X }_{+}$ represents the out-of-phase inter-well motion, in which the atoms and molecules oscillate inversely, whereas $\tilde{ X }_{-}$ represents the in-phase inter-well motion, in which the atoms and molecules oscillate in the same direction. For the parameters defined in Sec. \ref{sec:model}, $\omega_{+}$($\omega_{-}$) represents to the out-of-phase (in-phase) mode. In more general, the eigen vectors of $\omega_{\pm}$ modes are the linear combination of in-phase and out-of-phase motions. In addition, the corresponding frequencies $\omega_{\pm}$ are reduced to the Josephson frequencies in a single-component case in the limit of $g \rightarrow 0$ and $U_{ab} \rightarrow 0$.

\section{Particle-Localized ground states induced by internal Atom-molecule tunneling}

\label{sec:symmetry-breaking_ground_states}
When the atom-molecule interaction $U_{ab}$ and the atom-molecule internal tunneling $g$ are small, the ground state is symmetric, i.e. the particle populations in two wells are equal to each other. In this section, we will show that the atom-molecule internal tunneling induces the particle-localized ground state. This transition from the non-localized ground state to the localized one is signaled by the dynamical instability of the in-phase mode. In what follows, we investigate the dynamical stability of the original symmetric ground state, by using the excitation spectra derived in the previous section. After the stability analysis, we investigate the stationary states and show the bifurcation of the symmetric stationary state to the asymmetric ones. The cause of particle-localization will be discussed in Sec. \ref{sec:effective_attractive} and \ref{sec:stability}. In Sec. \ref{sec:stability} and \ref{sec:re_tra}, we also discuss the possibility of the reentrant transition.

Here, we explain the parameters of the particle interactions. We set the atomic interaction strength $U_{a}$ and the molecular interaction strength $U_{b}$ as described in Sec. \ref{sec:model} throughout this section. In what follows, we set the atom-molecule interaction strength as $U_{ab} / J_{a} = - 2.3 \times 10^{-2}$ (except the stability diagram Fig. \ref{fig:phase_graph} and \ref{fig:reentrant} varying $U_{ab}$ and $g$), whose absolute value is slightly smaller than $U_{a}$.

\subsection{Particle localization transition}
\label{sec:localization}

\begin{figure}[htbp]
\includegraphics[width=8.3cm]{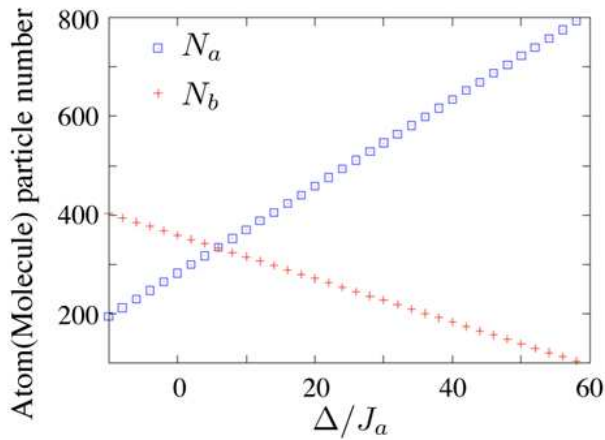}% Here is how to import EPS art
\caption{$\Delta$-dependence of particle populations at symmetric stationary states. $\Box$($+$) represents $N_{a}\left(N_{b}\right)$.}
\label{fig:fig_1}
\end{figure}

First, we determine the atom-molecule energy difference $\Delta$, which we use in this section. The stationary states are determined by solving Eq. (\ref{eq:stationary_1})-(\ref{eq:stationary_4}). In the absence of the atom-molecule internal tunneling there only exists the symmetric stationary state, which is the ground state. The $\Delta$-dependence of particle populations in the limit $g \rightarrow 0$ is shown in Fig. \ref{fig:fig_1}. From Fig. \ref{fig:fig_1} each condensate has a few hundreds of particles so that classical analysis introduced in Sec. \ref{sec:model_approximation} is appropriate. Hereafter we choose the atom-molecule energy difference as $\Delta / J_{a} = 3$ in order that the atomic and molecular particle numbers are almost the same.

\begin{figure}[htbp]
\includegraphics[width=8.3cm]{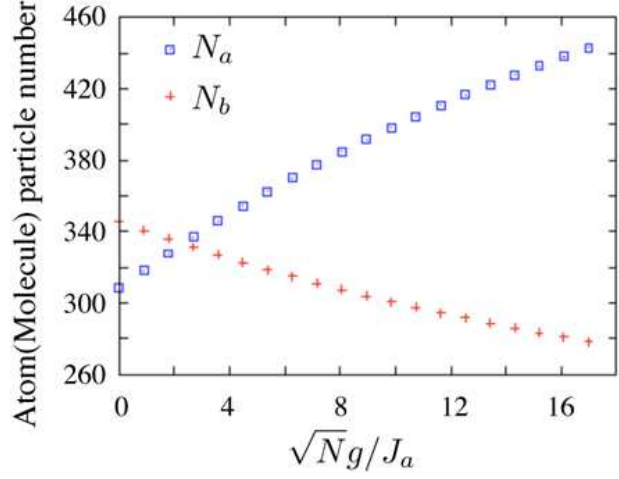}% Here is how to import EPS art
\caption{$g$-dependence of particle populations in the symmetric stationary states. $\Box$($+$) represents $N_{a}\left(N_{b}\right)$.}
\label{fig:fig_2}
\end{figure}

\begin{figure}[htbp]
\includegraphics[width=8.3cm]{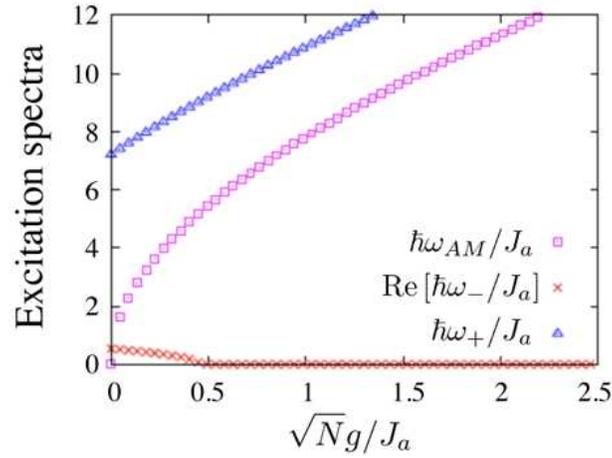}% Here is how to import EPS art
\caption{$g$-dependence of excitation spectra of the symmetric stationary states.}
\label{fig:fig_3}
\end{figure}

\begin{figure}[htbp]
\includegraphics[width=8.3cm]{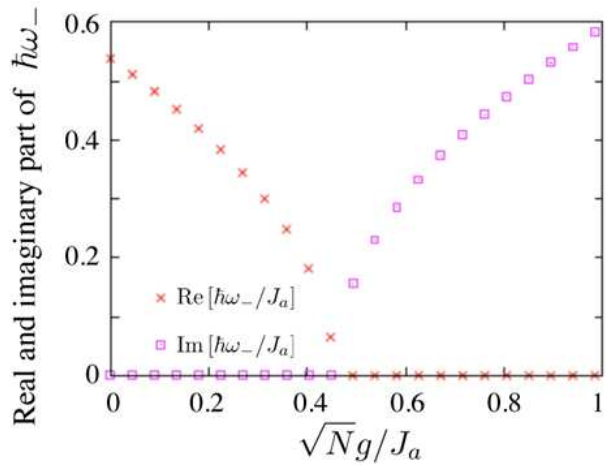}% Here is how to import EPS art
\caption{$g$-dependence of imaginary part and real part of $\hbar \omega_{-}$ at symmetric stationary states.}
\label{fig:fig_4}
\end{figure}

In Fig. \ref{fig:fig_2}, we investigate the $g$-dependence of particle populations at the symmetric stationary state. From this figure we conclude that the particle populations are large enough for applying the mean-field approximation in a wide range of atom-molecule internal tunneling $g$. We also find that the atomic populations in the ground states grows, by increasing the atom-molecule internal tunneling strength. In the symmetric stationary state, where $2 \theta_{aL(aR)} - \theta_{bL(bR)} = 0$, $N_{a} = N_{aL(aR)}$ and $N_{b} = N_{bL(bR)}$, the internal tunneling term in the Hamiltonian (\ref{eq:classical_Hamiltonian}) is reduced to be $- 4 g N_{a} \sqrt{ N_{b} }$. Because the order of $N_{a}$ is larger than that of $N_{b}$, the symmetric stationary state tends to lower the total energy by increasing $N_{a}$ rather than $N_{b}$ in the large $g$ region. 
\begin{figure}[htbp]
\includegraphics[width=8.3cm]{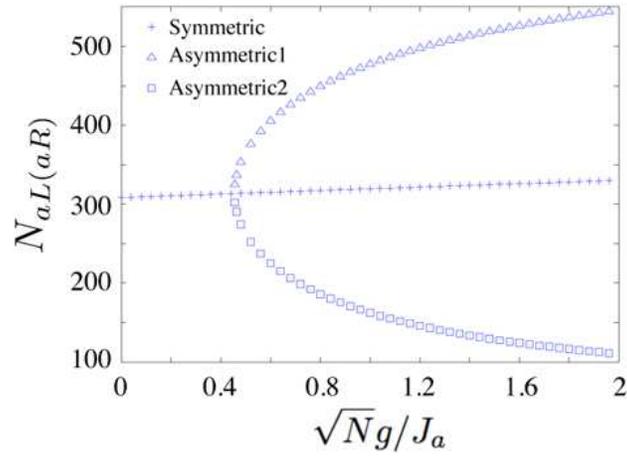}% Here is how to import EPS art
\caption{$g$-dependence of atomic particle populations at symmetric(+ points) and asymmetric stationary states($\bigtriangleup$ and $\Box$ points).}
\label{fig:fig_5}
\end{figure}

\begin{figure}[htbp]
\includegraphics[width=8.3cm]{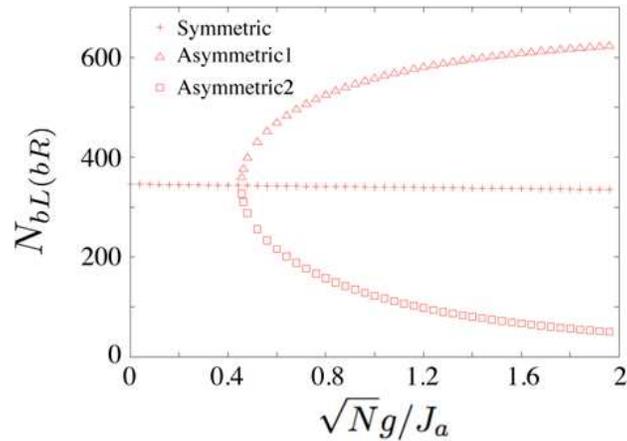}% Here is how to import EPS art
\caption{$g$-dependence of molecular particle populations at symmetric(+ points) and asymmetric stationary states($\bigtriangleup$ and $\Box$ points).}
\label{fig:fig_6}
\end{figure}

Next, we investigate the dynamical stability of the symmetric stationary state by looking at the excitation frequencies. Fig. \ref{fig:fig_3} is the excitation spectra. $\omega_{+}$ and $\omega_{AM}$ are always real and close to each other in the large $g$ region. On the other hand, as shown in Fig. \ref{fig:fig_3}, $\omega_{-}$ goes to zero at the finite atom-molecule internal tunneling $g$. As shown in Fig. \ref{fig:fig_4}, the imaginary part of $\omega_{-}$ emerges at the same value of $g$, while the other modes are still dynamically stable. This fact indicates the occurrence of symmetry-breaking phase transition.
\begin{figure}
%\begin{minipage}{9pc}
\includegraphics[width=8.3cm]{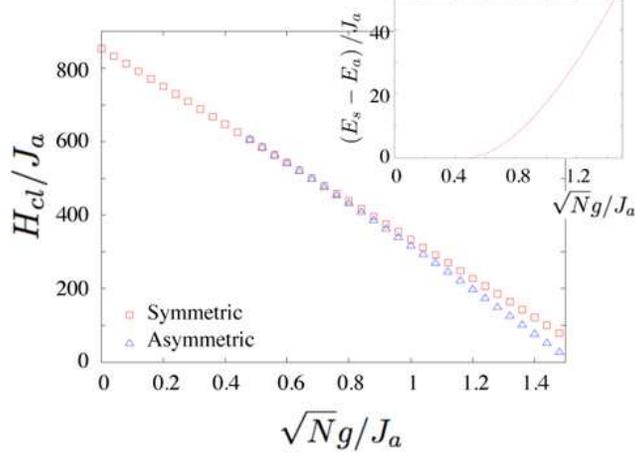}% Here is how to import EPS art
\caption{$g$-dependence of total energy at symmetric and asymmetric stationary states. The inset is energy difference between symmetric and asymmetric stationary states. $E_{s} \left( E_{a} \right)$ represents the total energy of symmetric(asymmetric) stationary state.}
\label{fig:fig_7}
\end{figure}
%\end{minipage}\hspace{0.5pc}
%\begin{minipage}{9pc}
\begin{figure}
\includegraphics[width=8.3cm]{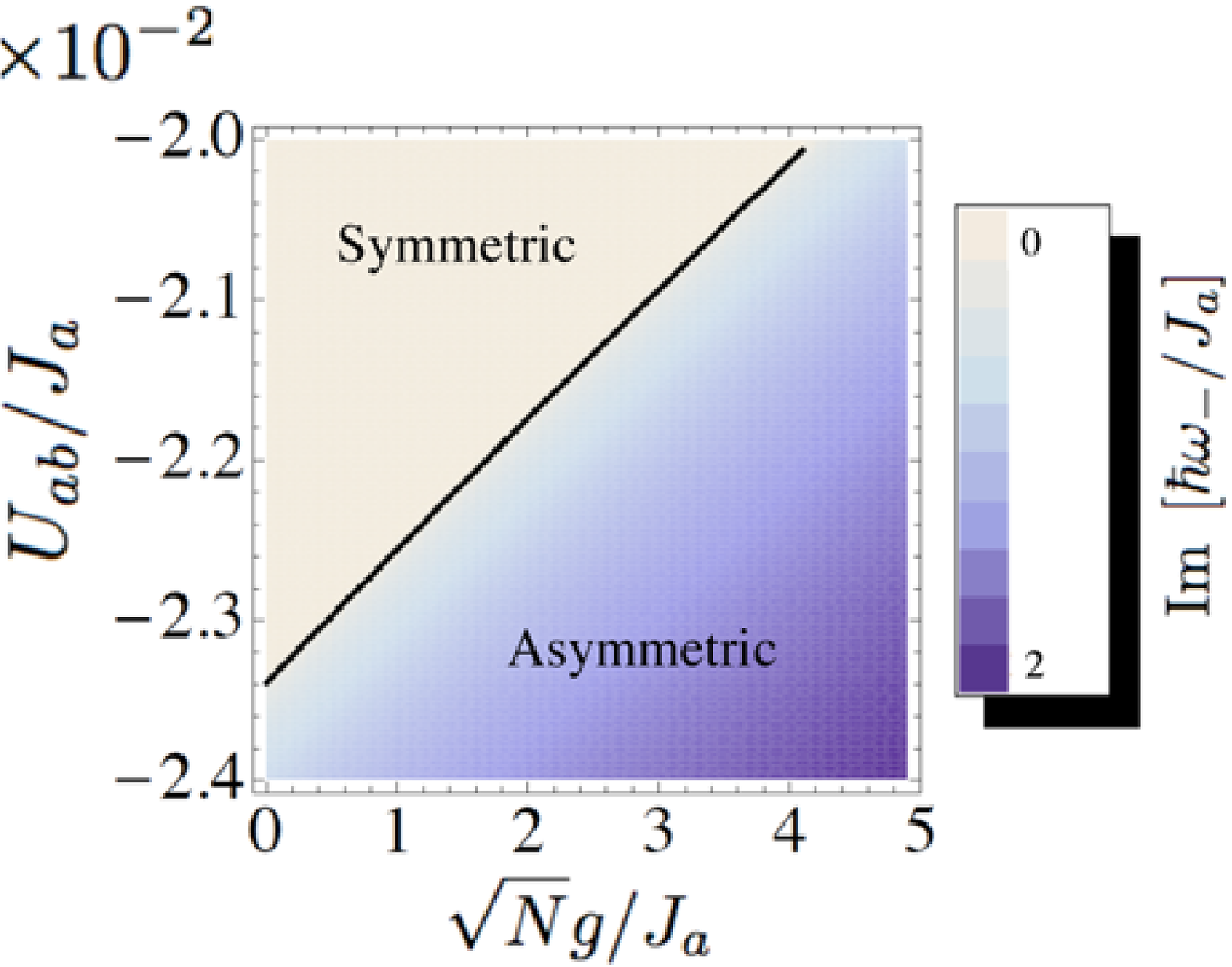}% Here is how to import EPS art
\caption{The stability phase diagram for the symmetric stationary state at $\Delta = 3 J_{a}$. The black line represents $\hbar \omega_{-} = 0$.}
\label{fig:phase_graph}
%\end{minipage}
\end{figure}

In order to confirm the appearance of the particle-localized ground state, we investigate the particle populations in the stationary state by solving Eqs. (\ref{eq:stationary_1})-(\ref{eq:stationary_4}). Figs. \ref{fig:fig_5} and \ref{fig:fig_6} show the atomic and molecular populations in the stationary solutions. These represent the bifurcation of the populations, which means the appearance of the new stationary states induced by the atom-molecule internal tunneling. In these new stationary states, the populations of atoms and molecules are localized in the same well, which means the  breaking of the left-right symmetry of a double-well potential. The value of $g$ at the bifurcation point is the same as that of the point, where dynamical instability occurs. Furthermore, we compare the total energies of the symmetric and asymmetric ground states. Fig. \ref{fig:fig_7} shows that the total energy of asymmetric state is lower than that of the symmetric state, which means the alteration of the ground state.

Finally, we investigate the stability of symmetric stationary states by varying $U_{ab}$ and $g$. The stability diagram of the symmetric stationary state is presented in Fig. \ref{fig:phase_graph}. From this figure, the large $g$ and negative $U_{ab}$ induce the asymmetric ground state. 

\subsection{The effective attractive interaction}
\label{sec:effective_attractive}
The effective inter-atomic attractive interaction mediated by molecular bosons is often discussed in the context of BCS-BEC crossover in a Fermi gas with Feshbach resonance using the two-channel model\cite{Ohashi_Griffin_PRL}. We discuss the same type of attractive interaction in the four mode model, assuming the simplest case $J_{b} = U_{b} = U_{ab} = 0$. 

In this simple case, we can solve Eqs. (\ref{eq:stationary_3}) and (\ref{eq:stationary_4}) for $\sqrt{N_{bL(bR)}}$ as, 
\begin{eqnarray}
\sqrt{ N_{bL(bR)} } = g \frac{ N_{aL(aR)} }{ \Delta - 2 \mu }.
\label{eq:from_Six}
\end{eqnarray}
By using this in the grand canonical energy, we obtain
\begin{eqnarray}
K = - 2 J_{a} \sqrt{ N_{aL} N_{aR} } + \frac{1}{2} \tilde{ U }_{a} \left( N_{aL}^2 + N_{aR}^2 \right) - \mu \left( N_{aL} + N_{aR} \right),
\label{eq:eff_grand}
\end{eqnarray}
where
\begin{eqnarray}
\tilde{ U }_{a} \equiv U_{a} - \frac{ 2 g^2 }{ \Delta - 2 \mu }.
\end{eqnarray}
This clearly shows that the effect of the atom-molecule tunneling modifies the inter-atomic interaction $U_{a}$ to the effective interaction $\tilde{U}_{a}$. From Eq. (\ref{eq:from_Six}), it is clear that $\Delta > 2 \mu$ always. Therefore, {\it the contribution from the atom-molecule tunneling is always attractive in the case $J_{b} = U_{b} = U_{ab} = 0$}.  

In order to investigate the transition from the symmetric to asymmetric state, we write
\begin{eqnarray}
N_{aL} = N_{a} \left( 1 + x \right), \quad N_{aR} = N_{a} \left( 1 - x \right),
\end{eqnarray}
and expand $K$ in $x$. We obtain
\begin{eqnarray}
K \simeq - 2 N_{a} J_{a} + N_{a}^{2} \tilde{U}_{a} - 2 \mu N_{a} + N_{a} \left( J_{a} + N_{a} \tilde{U}_{a} \right) x^2 + \frac{ N_{a} J_{a} }{ 4 } x^4. 
\end{eqnarray}
This clearly shows that the ground state is determined by the competition between the inter-well tunneling $J_{a}$ and the effective interaction $\tilde{U}_{a}$. The transition from the symmetric solution $\left( x = 0 \right)$ to the asymmetric solution $\left( x \ne 0 \right)$ occurs when the sign of the coefficient of the quadratic term changes from positive to negative. More explicitly, the asymmetric state become the ground state when $J_{a} + N_{a} \tilde{U}_{a} < 0$. By using (\ref{eq:from_Six}), this condition is written as 
\begin{eqnarray}
J_{a} + N_{a} U_{a} < 2 g \sqrt{ N_{b} }.
\label{eq:attractive_condition}
\end{eqnarray}
From this condition, the atom-molecule tunneling always tend to make the asymmetric ground state. This analysis assumed the simple case $J_{b} = 0, U_{b} = 0, U_{a} = 0$. One might expect that the quantitative results do not change for the general case, as long as $J_{b}, U_{b}$ and $U_{ab}$ are small. However, in the following sections, we show that the atom-molecule tunneling has also the effect creating the symmetric ground state in the general case $J_{b} \ne 0, U_{b} \ne 0, U_{ab} \ne 0$.

\subsection{The stability condition of the symmetric stationary state}
\label{sec:stability}
In this section, we consider the general case $J_{b} \ne 0, U_{b} \ne 0, U_{ab} \ne 0$. In this case, we have not been able to reduce the ground canonical energy in a simple closed form in terms atomic populations $N_{aL}$ and $N_{aR}$. Thus, we use the excitation spectra in order to discuss the stability of the symmetric state.

From Eq. (\ref{eq:spectra_new}), the dynamical instability condition $\left( \hbar \omega_{-} \right)^2 < 0$ is reduced to be
\begin{eqnarray}
\left( \frac{ J_{a} }{ N_{a} } + U_{a} \right) \left(  \frac{ J_{b} }{ N_{b} } + U_{b} + \frac{ g N_{a} }{  2 N_{b} \sqrt{ N_{b} } } \right) < \left( - \frac{ g }{ \sqrt{ N_{b} } } + U_{ab}  \right)^2.
\label{eq:sta_con}
\end{eqnarray}
When this inequality is satisfied, $\hbar \omega_{-}$ mode becomes dynamically unstable, signifying that the ground state becomes asymmetric. This condition generalizes Eq. (\ref{eq:attractive_condition}) to including $J_{b}, U_{b}$ and $U_{ab}$. The remarkable difference from (\ref{eq:attractive_condition}) is that the atom-molecule tunneling strength $g$ appears in both sides of the equation. This means that the atom-molecule internal tunneling does not only tend to create the asymmetric ground state but also symmetric one. This non-monotonic behavior has not been found in the simplest model neglecting $J_{b}, U_{b}, U_{ab}$.

 The term including $g$ is $g N_{a} / \left( 2 N_{b} \sqrt{N_{b}} \right)$ in the left hand side, and $g / \sqrt{ N_{b} }$ in the right hand side. The ratio of these terms is $N_{a} / \left( 2 N_{b} \right)$, which is the ratio of the numbers of atomic-state and molecular-state particles. This is controlled by the atom-molecule energy difference $\Delta$ like Fig. \ref{fig:fig_1} as explained in sec. \ref{sec:localization}. 

The large $\Delta$ tends to increase $N_{a}$ as seen from Fig. \ref{fig:fig_1}. Therefore, when $\Delta \left( > 0 \right)$ is large, by increasing $g$ in Eq. (\ref{eq:sta_con}), the term including $g$ in the left hand side is more enlarged than the one in the right hand side. In contrast, when $\Delta$ is small, the term including $g$ in the left hand side does not have a significant role in the symmetric-asymmetric transition. Therefore, the atom-molecule tunneling tends to create asymmetric ground state as Eq. (\ref{eq:attractive_condition}).

However, this $\Delta$-dependence is not the case in the large $g$ region. Since the right hand side is quadratic in $g$, the right hand side of Eq. (\ref{eq:sta_con}) is enlarged by $g$ independently to the atom-molecule energy difference $\Delta$ in the large $g$ region. In this large $g$ region, Eq. (\ref{eq:sta_con}) reduces to Eq. (\ref{eq:attractive_condition}). 

Therefore, when $\Delta \left( > 0 \right)$ is large, the atom-molecule tunneling has the tendency to create the symmetric ground state in the small $g$ region, whereas it tends to create the asymmetric ground state in the large $g$ region. The influence of the atom-molecule tunneling on the ground state depends on the strength of the atom-molecule tunneling when $\Delta \left( > 0 \right)$ is large. In Sec \ref{sec:localization}, we investigated the small $\Delta$ region. In the next section, we investigate the large $\Delta$ region.

\subsection{Reentrant transition}
\label{sec:re_tra}
In this section, we discuss the possibility of the reentrant transition, where the ground state changes in the order of asymmetric-symmetric-asymmetric by increasing $g$. The reentrant transition occurs when the positive $\Delta$ and the negative $U_{ab}$ are large. As explained in the previous section, when $\Delta \left( > 0 \right)$ is large, the atom-molecule tunneling tends to create the symmetric ground state in the small $g$ region. Therefore, when the negative $U_{ab}$ is large enough to create the asymmetric ground state at $g = 0$, the finite $g$ can cause the transition to the symmetric ground state from the asymmetric one. In contrast, in the large $g$ region, the atom-molecule tunneling tends to create the asymmetric ground state, and the transition from symmetric to asymmetric ground state is possible. In this way, by increasing the internal tunneling $g$ at the sufficiently large $\Delta$, the ground state turns from the asymmetric state to the symmetric state, and again, goes to the asymmetric phase. 
\begin{figure}
\includegraphics[width=8.3cm]{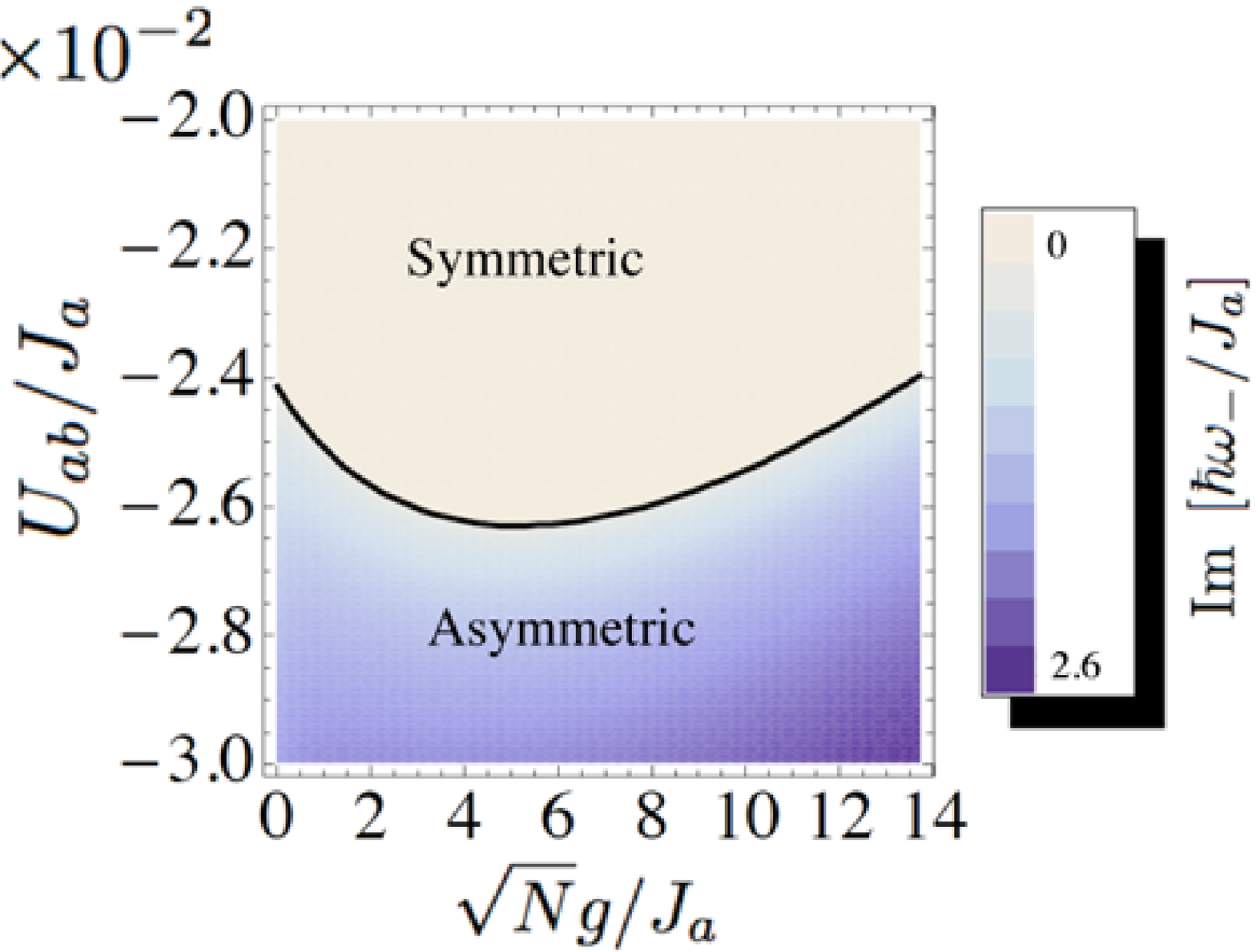}% Here is how to import EPS art
\caption{The stability phase diagram for the symmetric stationary state at $\Delta = 50 J_{a}$. The black line represents $\hbar \omega_{-} = 0$.}
\label{fig:reentrant}
\end{figure}

In fact, the large $\Delta$ changes the stability phase diagram as Fig. \ref{fig:reentrant}, which is quite different from Fig. \ref{fig:phase_graph}. In this figure, around $U_{ab} / J_{a} = - 2.5 \times 10^{-2}$ at $g=0$, the symmetric stationary state is unstable, and increasing the internal tunneling $g$, the symmetric state changes to be stable. Increasing $g$ further, the symmetric state turns to be unstable again. We note that asymmetric stationary states exist, when the symmetric sate is unstable. In this region, the asymmetric states are the ground states.

In what follows, we estimate the value of $\Delta$ required to cause the reentrant transition at the ground state. From Eq. (\ref{eq:sta_con}), the phase boundary between the stable and unstable phases is given by 
\begin{eqnarray}
 U_{ab} =\frac{ g }{ \sqrt{ N_{b} } } - \sqrt{ \left( \frac{ J_{a} }{ N_{a} } + U_{a} \right) \left( \frac{ J_{b} }{ N_{b} } + U_{b} + \frac{ g N_{a} }{ 2 N_{b} \sqrt{ N_{b} } } \right) }.
\end{eqnarray}
If the reentrant transition does not occur, the stability phase diagram appears as Fig. \ref{fig:phase_graph}, where the phase boundary is almost straight line. On the contrary, in the case that the reentrant transition occurs as in Fig. \ref{fig:reentrant}, the phase boundary line has the minimal $U_{ab}$ value. Therefore, from the condition $\partial U_{ab} / \partial g = 0$, we can derive the condition for the reentrant transition. We find that
\begin{eqnarray}
g_{min} = \frac{ 1 }{ 8 \sqrt{ N_{b} } } \left( J_{a} + N_{a} U_{a} \right) - \frac{ 2 \sqrt{ N_{b} } }{ N_{a} } \left( J_{b} + N_{b} U_{b} \right),
\end{eqnarray}
where we define $g_{min}$ as $g$ at the minimal $U_{ab}$ point. The condition for the reentrant transition is given by $g_{min} > 0$. In the weak coupling limit $N_{a}U_{a} \ll J_{a}$ and $N_{b}U_{b} \ll J_{b}$, the condition $g_{min} > 0$ is reduced to be $N_{a} / N_{b} > 16 J_{b} / J_{a}$, and in the strong coupling limit $N_{a}U_{a} \gg J_{a}$ and $N_{b}U_{b} \gg J_{b}$, the condition for the reentrant transition is $N_{a} / N_{b} > 4 \sqrt{ U_{b} / U_{a} }$. In this paper, we take $N U_{a} / J_{a} = N U_{b} / J_{b} = 60$ following the discussion in Sec. \ref{sec:model}. Therefore, we are in the strong coupling limit. Using the parameters in Sec. \ref{sec:model}, $U_{b} / U_{a} = 1/2$, and $ N_{a} / N_{b} > 2 \sqrt{ 2 }$.
\begin{figure}
\includegraphics[width=8.3cm]{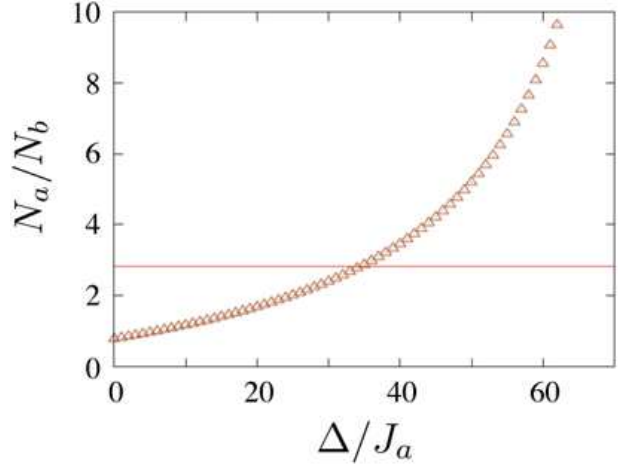}% Here is how to import EPS art
\caption{$\Delta$-dependence of $N_{a}/N_{b}$ at the symmetric stationary state(triangle points). The solid line represents the $N_{a}/N_{b}=2 \sqrt{2}$.}
\label{fig:delta_NaNb}
\end{figure}
Using this criterion, $\Delta \ge 40 J_{a}$ is needed to cause the reentrant transition from Fig. \ref{fig:delta_NaNb}. 

We now examine the experimental possibility realizing this large detuning $\Delta$. First, we estimate the value of $J_{a} / \hbar$. Within a two mode model, the Josephson frequency in the single component case\cite{Raghavan} is $\hbar \omega_{J} = 2 J_{a} \sqrt{ 1 + \Lambda }$. In the single BJJ experiment\cite{BJJ}, the Josephson frequency is estimated as $\omega_{J} = 40 \left( \textrm{ms} \right)^{-1}$, and $\Lambda = 15$ as explained in Sec. \ref{sec:model}. Using these values, we find $J_{a} / \hbar \simeq 20 \textrm{s}^{ - 1 }.$ Using this value, we estimate the value of $\Delta = 50 J_{a}$, which we use in Fig. \ref{fig:reentrant}. This gives
\begin{eqnarray}
\frac{ \Delta }{ \hbar } = 50 \times 20 \textrm{s}^{ - 1 } = 1 {\textrm k}{\textrm H}{\textrm z}.
\label{eq:Delta}
\end{eqnarray}
We compare this value with the experimental value of Feshbach resonance using $^{87}\textrm{Rb}$\cite{AMRabi}. In this experiment, the energy difference between the open-channel(closed-channel), which corresponds to the atomic(molecular) state, is $\Delta \mu \left( B - B_{res} \right)$, where $\Delta \mu$ is the difference of magnetic moments between these states, and $B$ is the strength of the magnetic field. $B_{res}$ is the strength of the magnetic field at the Feshbach resonance point. In this experiment, $\Delta \mu$ is estimated as $\Delta \mu = 2 \pi \hbar \times 111 \textrm{ kHz/G }$. In this experiment, $B$ ends typically $50 \textrm{mG}$ away from the Feshbach resonance. Using this value, $\Delta \mu \times \left( B - B_{res} \right) / \hbar \simeq 2 \pi \times 5 {\textrm k}{\textrm H}{\textrm z}$. This is larger than the estimated value given in (\ref{eq:Delta}).

\section{Summary}
\label{sec:sum}

We investigated how the atom-molecule internal tunneling changes the ground state of atom-molecule BECs in a double-well potential. From the linear stability analysis, we showed that  the atom-molecule internal tunneling induces the particle-localized ground state through dynamical instability. This is quite different from the single component BJJ because the tunneling terms between BECs tends to prevent localizations in a single component BJJ\cite{Trippenbach_QPT_BJJ}.  

As explained in Sec. \ref{sec:stability} the particle localization is caused by the fact that the inter-atom-molecule tunneling behaves like inter-atomic {\it attractive} interaction effectively. We showed that this effective interaction is always attractive in the absence of $J_{b}, U_{b}$ and $U_{ab}$. This is the same type effective interaction with that in the two-channel model often used in the study of BEC-BCS crossover in a Fermi gas with Feshbach resonance\cite{Ohashi_Griffin_PRL}. 

However, the situation can be quite different in the general case that $J_{b} \ne 0, U_{b} \ne 0$ and $U_{ab} \ne 0$. We pointed out the possibility of the reentrant transition, which cannot be understood in terms of the effective attractive interaction. In the large $\Delta \left( > 0 \right)$ region, the ground state changes from localized to non-localized, and again to localized, with increasing the atom-molecule internal tunneling. From this result, we pointed out the possibility that the atom-molecule internal tunneling exhibits the rich physics undiscovered in the treatment of the simple two-channel model neglecting the atom-molecule interaction and inter-molecule interaction.  

Next, we point out that quantum fluctuations do not prevent the particle localization. In order to consider the effect of quantum fluctuations, we performed the exact diagonalization of the Hamiltonian (\ref{eq:am_H})\cite{Motohashi_P2}. From this full-quantum treatment, we conclude that the ground state becomes the superposition of the particle-localized states in the left and right well.

Finally, we note that the reentrant transition cannot occur in a binary BEC mixture. By replacing the internal tunneling term in the (\ref{eq:am_H}) with $\left( \hat{b}_{L}^{\dag} \hat{a}_{L} + \hat{b}_{L} \hat{a}^{\dag}_{L} + \hat{b}_{R}^{\dag} \hat{a}_{R} + \hat{b}_{R} \hat{a}^{\dag}_{R} \right)$, we have the Hamiltonian for a binary BEC mixture. From the same procedure as in Sec. \ref{sec:model}, we can obtain the equation for a binary BEC mixture similar to Eq. (\ref{eq:sta_con}). From this equation, we can conclude that the reentrant transition is unique to atom-molecule mixture BECs.

\begin{acknowledgments}
The authors would like to thank S. Konabe, S. Watabe, T. Miyakawa, S. Tsuchiya, I. Danshita, T. Kato, T. Yamamoto, T. Ozaki, J. Haruyama and Y. Iwami for valuable comments and discussions.
\end{acknowledgments}

\appendix

\section{Dynamical instability and second order phase transition}
\label{sec:effective_potential}
In this Appendix, we briefly review the general relation between symmetry breaking and dynamical instability. We suppose the classical Hamiltonian having $f$ degrees of freedom as $H = \sum^{f}_{i=1} \frac{p^2_i}{2} + V_{\it eff} \left( x_{1} , x_{2} , \cdots . x_{f} \right)$, where $x_{i} ( i = 1, ... , f )$ are the dynamical variables and $p_{i} ( i = 1, ... , f )$ are their canonical momentums. The coefficients of kinetic terms have been normalized by scaling the dynamical variables. Expanding this effective potential around stationary states to second order in the fluctuations, one obtains $H \simeq \sum_{i}^{f} \frac{p^2_i}{2} + \frac{1}{2} \sum_{i , j} \frac{ \partial^2 H }{ \partial x_{i} \partial x_{j} } \delta x_{i} \delta x_{j} \equiv \sum_{i}^{f} \frac{p^2_i}{2} + \frac{1}{2} \mathbf{\delta x}^{\dag} \mathbf{ V }  \mathbf{\delta x}$. Here we defined $\mathbf{\delta x}^{\dag} \equiv \left( \delta x_{1} , \delta x_{2} , \cdots \delta x_{f} \right)$ and $\mathbf{ V }_{ i , j } \equiv { \partial^2 H } / { \partial x_{i} \partial x_{j} }$. The linearized Hamilton equations are given by, $\dot{ p }_{i} = - \frac{ \partial H }{ \partial x_{i} } = - \sum_{j=1}^{f} \frac{ \partial^2 H }{ \partial x_{i} \partial x_{j} } \delta x_{j}, \delta \dot{ x }_{i} = \frac{ \partial H }{ \partial p_{i} } = p_{i}$, then $\delta \ddot{ x }_{i} = - \sum_{j=1}^{f} \frac{ \partial^2 H }{ \partial x_{i} \partial x_{j} } \delta x_{j} \Leftrightarrow \delta \ddot{ \mathbf{ x } } = - \mathbf{ V } \delta \mathbf{ x }$. Defining the eigen frequency as $\omega$, $x_{i} \propto e^{ \pm i \omega t} , ( i = 1,...,f )$, we find that the eigen value of $\mathbf{V}$ is eqaul to $\omega^2$. 

It is thus clear that the matrix $\mathbf{V}$ is positive definite, if we expand around the ground state. Then, if any of eigen values change to be minus at a ground state by varying parameters, this state is no longer a ground state. In general, the original ground state becomes a saddle point or a local maximum, and the new ground states are bifurcated around the original one. 

Consequently, the second order phase transition of $V_{\it eff}$ corresponds to the fact that an excitation spectrum $\hbar \omega$ becomes purely imaginary number. A purely imaginary $\omega$ represents a slipping off from the original ground state to the symmetry-breaking ones. Since $\mathbf{V}$ is the symmetric matrix, the eigenvalue is always real, and thus in this second order phase transition of a ground state, an excitation spectra becomes purely imaginary number.

\section{Parameters}
\label{sec:parameter_appendix}
The parameters are defined as follows :
\begin{eqnarray}
&& J_{i} \equiv - \int d \mathbf{r} \Phi_{iL}^{*} \left[ - \frac{\hbar^2}{ 2m_{i} } \nabla^{2} + V_{ \rm{ext} } ( \mathbf{r} ) \right] \Phi_{iR} ,
\label{eq:p_1} \\
&& E_{i}^{0} \equiv  \int d \mathbf{r} \Phi_{iL}^{*} \left[ - \frac{\hbar^2}{ 2m_{i} } \nabla^{2} + V_{ \rm{ext} } ( \mathbf{r} ) \right] \Phi_{iL} =  \int d \mathbf{r} \Phi_{iR}^{*} \left[ - \frac{\hbar^2}{ 2m_{i} } \nabla^{2} + V_{ \rm{ext} } ( \mathbf{r} ) \right] \Phi_{iR} ,
\label{eq:p_2} \\
&& U_{i} \equiv g_{i} \int d \mathbf{r} | \Phi_{iL} |^{4} = g_{i} \int d \mathbf{r} | \Phi_{iR} |^{4} ,
\label{eq:p_3} \\
&& U_{ab} \equiv g_{ab} \int d \mathbf{r} | \Phi_{aR} |^{2} | \Phi_{bR} |^{2} = g_{ab} \int d \mathbf{r} | \Phi_{aL} |^{2} | \Phi_{bL} |^{2} ,
\label{eq:p_4} \\
&& g \equiv \lambda \int d \mathbf{r} \Phi_{bL}^{*} \Phi_{aL} \Phi_{aL} = \lambda \int d \mathbf{r} \Phi_{bR}^{*} \Phi_{aR} \Phi_{aR} ,
\label{eq:p_5} \\
&& \Delta \equiv \delta \int d \mathbf{r} | \Phi_{bL} |^{2} + E_{b}^{0} - 2 E_{a}^{0} = \delta \int d \mathbf{r} | \Phi_{bR} |^{2} + E_{b}^{0} - 2 E_{a}^{0} ,
\label{eq:p_6}
\end{eqnarray}
where $i = a(b)$ represents atomic(molecular) BEC modes, and L(R) expresses the left(right) well respectively. 

\section{time-evolution equations}
\label{sec:app_1_1}
The exlicit forms of the Hamilton equations (\ref{eq:canonical_1})-(\ref{eq:canonical_4}) are as follows :
\begin{eqnarray}
\hbar \dot{N}_{aL} &=& - 2J_{a} \sqrt{ N_{aL} N_{aR} } \sin( \theta_{aR} - \theta_{aL} ) + 4 g N_{aL} \sqrt{ N_{bL} } \sin( 2 \theta_{aL} - \theta_{bL} ) \label{eq:te_1} \\
\hbar \dot{N}_{aR} &=& - 2J_{a} \sqrt{ N_{aL} N_{aR} } \sin( \theta_{aL} - \theta_{aR} ) + 4 g N_{aR} \sqrt{ N_{bR} } \sin( 2 \theta_{aR} - \theta_{bR} ) \label{eq:te_3} \\
\hbar \dot{N}_{bL} &=& - 2 J_{b} \sqrt{N_{bL} N_{bR} } \sin( \theta_{bR} - \theta_{bL} ) - 2 g N_{aL} \sqrt{ N_{bL} } \sin( 2 \theta_{aL} - \theta_{bL} ) \label{eq:te_5} \\
\hbar \dot{N}_{bR} &=& - 2 J_{b} \sqrt{N_{bL} N_{bR} } \sin( \theta_{bL} - \theta_{bR} ) - 2 g N_{aR} \sqrt{ N_{bR} } \sin( 2 \theta_{aR} - \theta_{bR} ) \label{eq:te_7} \\
\hbar \dot{ \theta }_{aL} &=& J_{a} \sqrt{ \frac{ N_{aR} }{ N_{aL} } } \cos( \theta_{aR} - \theta_{aL} ) - N_{aL} U_{a} - N_{bL} U_{ab} + 2 g \sqrt{N_{bL} } \cos( 2 \theta_{aL} - \theta_{bL} ) \label{eq:te_2} \\
\hbar \dot{ \theta }_{aR} &=& J_{a} \sqrt{ \frac{ N_{aL} }{ N_{aR} } } \cos( \theta_{aL} - \theta_{aR} ) - N_{aR} U_{a} - N_{bR} U_{ab} + 2 g \sqrt{N_{bR} } \cos( \theta_{bR} - 2 \theta_{aR} ) \label{eq:te_4} \\
\hbar \dot{ \theta }_{bL} &=& J_{b} \sqrt{ \frac{ N_{bR} }{ N_{bL} } } \cos ( \theta_{bR} - \theta_{bL} ) - \Delta - N_{bL} U_{b} - N_{aL} U_{ab} + g \frac{ N_{aL} }{ \sqrt{ N_{bL} } } \cos( 2 \theta_{aL} - \theta_{bL} ) \label{eq:te_6} \\
\hbar \dot{ \theta }_{bR} &=& J_{b} \sqrt{ \frac{ N_{bL} }{ N_{bR} } } \cos ( \theta_{bL} - \theta_{bR} ) - \Delta - N_{bR} U_{b} - N_{aR} U_{ab} + g \frac{ N_{aR} }{ \sqrt{ N_{bR} } } \cos( 2 \theta_{a2} - \theta_{bR} ) \label{eq:te_8}
\end{eqnarray}

\section{Canonical Transformation}
\label{sec:canonical_variables}
The relative phases in the ground state are $\theta_{aL(bL)} - \theta_{aR(bR)} = 0$,$2 \theta_{aL(aR)} - \theta_{bL(bR)} = 0$. Expanding cosines around these zero phases as $\cos \theta \simeq 1 - \frac{1}{2} \theta^2$, we obtain the quadratic form in the phase variables. Therefore, the classical Hamiltonian can be expanded to second order in phase fluctuations around the symmetric stationary state. Furthermore, we diagonalize the Hamiltonian about the phase fluctuations. As a result, one of the eigen value is zero, and therefore, the number of degrees of freedom decreases by one. We avoid this problem by introducing a fictitious parameter $\alpha$, and in the last of the calculations, we set $\alpha \rightarrow 0$.  After these procedures, we obtain the Hamiltonian as 
\begin{eqnarray}
H_{cl} \simeq \alpha \phi_{0}^2 + \frac{ 1 }{ 2 } U \phi_{AM}^2 + \Omega_{+} Z_{+} \phi_{+}^2 + \Omega_{-} Z_{-} \phi_{-}^2 + V_{eff} \left( N_{aL}, N_{aR}, N_{bL}, N_{bR} \right),
\end{eqnarray}
where %$S \equiv N_{a} J_{a}, T \equiv N_{b} J_{b}, U \equiv g N_{a} \sqrt{ N_{b} }$, 
\begin{eqnarray}
\Omega_{\pm} &\equiv& N_{a} J_{a} + N_{b} J_{b} + \frac{ 5 }{ 2 } g N_{a} \sqrt{ N_{b} } \nonumber \\
&& \pm \frac{ 1 }{ 2 } \sqrt{ 4 ( N_{a} J_{a} - N_{b} J_{b} )^2 + 12 ( N_{a} J_{a} - N_{b} J_{b} ) g N_{a} \sqrt{ N_{b} } + 25 g^2 N_{a}^2 N_{b} },
\end{eqnarray}
and $Z_{\pm} \equiv \left( 2 \alpha_{\pm}^{2} + 2 \right)^{-1}$, where $\alpha_{\pm} \equiv \left( \Omega_{\pm} - 2 N_{b} J_{b} - g N_{a} \sqrt{ N_{b} } \right) / \left( 2 g N_{a} \sqrt{ N_{b} } \right)$. We defined the notations as $\phi_{0} \equiv \frac{1}{2} \theta_{aL} + \frac{1}{2} \theta_{aR} + \theta_{bL} + \theta_{bR}, \phi_{AM} \equiv - 2 \theta_{aL} - 2 \theta_{aR} + \theta_{bL} + \theta_{bR}$ and $\phi_{\pm} \equiv \alpha_{\pm} \theta_{aL} - \alpha_{\pm} \theta_{aR} - \theta_{bL} + \theta_{bR}$. $\phi_{0}$ represents the whole increase of phases keeping the relative phases constant, $\phi_{\pm}$ are the inter-well oscillation modes, and $\phi_{AM}$ is the oscillation between the atomic and molecular states. $V_{eff}$ is
\begin{eqnarray}
&& V_{\it eff} = - 2 J_{a} \sqrt{N_{aL}N_{aR}} - 2 J_{b} \sqrt{N_{bL}N_{bR}} + \Delta ( N_{bL} + N_{bR} ) + \frac{U_{a}}{2} ( N_{aL}^2 + N_{aR}^2 ) + \frac{U_{b}}{2} ( N_{bL}^2 + N_{bR}^2 ) \nonumber \\
&& \qquad \quad + U_{ab} ( N_{aL} N_{bL} + N_{aR} N_{bR} ) - 2g \big[ N_{aL} \sqrt{N_{bL}} + N_{aR} \sqrt{ N_{bR} } \big]. \label{eq:effective}
\end{eqnarray}
We denote the canonical conjugate variables with $\phi_{0}, \phi_{+}, \phi_{-}, \phi_{AM}$ as $X_{i} ( i = 0, +, -, AM )$. These are related with the particle numbers as
\begin{eqnarray}
\left(
\begin{array}{cccc}   
X_{0} \\ 
X_{+} \\ 
X_{-}\\ 
X_{AM}
 \end{array}
 \right)
= 
\left(
\begin{array}{cccc} 
\frac{1}{5} &\frac{1}{5}  &\frac{2}{5} & \frac{2}{5} \\  
\frac{1}{2} \xi& - \frac{1}{2} \xi & \frac{1}{2} \alpha_{-} \xi & - \frac{ 1 }{2} \alpha_{-} \xi  \\ 
- \frac{1}{2} \xi & \frac{1}{2} \xi  & - \frac{1}{2}  \alpha_{+} \xi & \frac{ 1 }{2} \alpha_{+} \xi \\ 
 - \frac{1}{5} & - \frac{1}{5}  & \frac{1}{10}  & \frac{1}{10}
\end{array} 
\right)
\left(
\begin{array}{cccc}   
N_{aL} \\ 
N_{aR} \\ 
N_{bL}\\ 
N_{bR}
\end{array}
\right) , \nonumber \\
\label{eq:ryusi_mode}
\end{eqnarray}
where $ \xi \equiv \left( \alpha_{+} - \alpha_{-} \right)^{- 1} $. $X_{0}$ represents the increase of whole particle number, $X_{\pm}$ is the inter well oscillation mode, and $X_{AM}$ is the oscillation between the atomic and molecular states. The poisson brackets are $\{ X_{i} , \phi_{j} \} = i \delta_{i , j}\left( i , j = 0, +, -, AM \right)$, and therefore these are canonical conjugate variables. We then rescale $\phi_{0}, \phi_{\pm}, \phi_{AM}$ as $\tilde{ \phi }_{0} \equiv \sqrt{ 2 \alpha } \phi_{0} , \tilde{ \phi }_{AM} \equiv \sqrt{ g N_{a} \sqrt{ N_{b} } } \phi_{AM}, \tilde{ \phi }_{+} \equiv \sqrt{ 2 \Omega_{+} Z_{+} } \phi_{+} , \tilde{ \phi }_{-} \equiv \sqrt{ 2 \Omega_{-} Z_{-} } \phi_{-}$. Corresponding canonical transformation for the coordinate variables are $\tilde{ X }_{0} \equiv X_{0} / \sqrt{ 2 \alpha } , \tilde{ X }_{AM} \equiv X_{AM} / \sqrt{ g N_{a} \sqrt{ N_{b} } }, \tilde{ X }_{+} \equiv X_{+} / \sqrt{ 2 \Omega_{+} Z_{+} }, \tilde{ X }_{-} \equiv X_{-} / \sqrt{ 2 \Omega_{-} Z_{-} }$. It is easy to see that the poisson brackets are maintained as $\{ X_{i} , \phi_{j} \}  = \{ \tilde{ X }_{i} , \tilde{ \phi }_{j} \} = \delta_{i , j},\left( i , j = 0, +, -, AM \right)$. Therefore the above transformation is the canonical transformation. As a result of this procedure, {\it the $\alpha$ dependence of the coefficient of $\phi_{0}$ move to that of $\tilde{ X }_{0}$}. By setting the fictitious parameter $\alpha \rightarrow 0$ and eliminating the degrees $\tilde{ X }_{0}$ relating to global phase rotation, the degrees of the inter-well and internal tunnelings are decoupled in $V_{eff}$. Then, we finally arrive at the Hamiltonian (\ref{eq:di_Ham}) and (\ref{eq:V_eff_a_0}). 

\section{Linearized Hamilton equations}
\label{sec:app_1}
The linerarized Hamilton equations are $\hbar \delta \dot{ \tilde{ X } }_{i} = \partial H_{cl} / \partial \tilde{ \phi }_{i} = \tilde{ \phi }_{i}$, where $i = 0, AM, \pm$, and
\begin{eqnarray}
\hbar \dot{ \tilde{ \phi } }_{0} &=& - \frac{ \partial V_{eff} }{ \partial \tilde{ X }_{0} } \nonumber \\
&=&  - \alpha \left( U_{a} + 4 U_{b}^{e} - 4 U_{ab}^{e} \right) \delta \tilde{ X }_{0} - \sqrt{ 2 \alpha C } \left( - 2 U_{a} + 2 U_{b}^{e} + 3 U_{ab}^{e} \right) \delta \tilde{ X }_{AM}. \label{eq:phase_linear_0} \\
\hbar \dot{ \tilde{ \phi } }_{AM} &=& - \frac{ \partial V_{eff} }{ \partial \tilde{ X }_{AM} } \nonumber \\
&=& - \sqrt{ 2 \alpha C } \left( - 2 U_{a} + 2 U_{b}^{e} + 3 U_{ab}^{e} \right) \delta \tilde{ X }_{0} - 2 C \left( 4 U_{a} + U_{b}^{e} + 4 U_{ab}^{e} \right) \delta \tilde{ X }_{AM} \label{eq:phase_linear_AM} \\
\hbar \dot{ \tilde{ \phi } }_{\pm} &=& - \frac{ \partial V_{eff} }{ \partial \tilde{ X }_{\pm} } \nonumber \\
&=& - 4 \Omega_{\pm} Z_{\pm} \left( \alpha_{\pm}^2 J_{a}^{e} + J_{b}^{e} + 2 \alpha_{\pm} U_{ab}^{e} \right) \delta \tilde{ X }_{\pm}  \nonumber \\
&& - 4 \sqrt{ \Omega_{+} \Omega_{-} Z_{+} Z_{-} } \left( - J_{a}^{e} + J_{b}^{e} + \left( \alpha_{+} + \alpha_{-} \right) U_{ab}^{e} \right) \delta \tilde{X}_{\mp}.
\label{eq:phase_linear_pm}
\end{eqnarray}
Setting the fictitious parameter $\alpha \rightarrow 0$, the contribution from the degrees $\tilde{ X }_{0}, \tilde{ \phi }_{0}$ relating to global phase rotation is eliminated. 

\bibliography{bunken}

%\begin{thebibliography}{99}
%\bibitem{1} M. Albiez, R. Gati, J. F\"{o}lling, S. Hunsmann, M. Cristiani and M. Oberthaler, Phys. Rev. Lett. {\bf 95} 010402(2005) 
%\end{thebibliography}

\end{document}